\documentclass[12pt]{article}
\begin{document}
\begin{flushright}
USM-TH-100
\end{flushright}

\begin{center}  {Searching for Extra Dimensions in the Early Universe}
\end{center}
\vspace{.2in}
\begin{center}
Douglas J. Buettner
\\Aerospace Corporation\\P.O. Box 92957  - M1/112\\
Los Angeles, CA  90009-2957
\footnote{e-mail address:
Douglas.J.Buettner@aero.org}
\end{center}
\vspace{.2in}
\begin{center}
P.D. Morley
\\General Dynamics Corporation\\National Systems Group\\14700 Lee Road
Chantilly, VA 20151\footnote{e-mail address:
peter.morley@veridian.com}
\end{center}
\vspace{.2in}
\begin{center}
Ivan Schmidt
\\Department of Physics \\ Universidad T\'{e}cnica Federico
Santa Mar\'\i a \\ Casilla 110-V, Valpara\'\i so, Chile\footnote{e-mail address:
ischmidt@fis.utfsm.cl}
\end{center}
\begin{abstract}
\baselineskip=0.7cm
We investigate extra spatial dimensions ($D =
3+\epsilon$) in the early universe using very high resolution molecular rotational spectroscopic
data derived from a large molecular cloud containing moderately cold carbon monoxide gas at Z $\approx 6.42$.
It turns out that the $\epsilon$-dependent quantum mechanical wavelength transitions are solvable for a linear molecule and we present the solution here. The CO microwave data allows a very precise
determination of $< \epsilon > = -0.00000657 \pm .10003032$. The probability that
$ < \epsilon > \neq 0$ is one in 7794, only 850 million years (using the standard cosmology) after the Big Bang.
\end{abstract}

Recently, microwave molecular CO spectroscopic transitions has
been detected [1,2] from the most distant quasar currently known,
SDSS J114816.64+525150.3, at a redshift Z $\approx$ 6.42. The data
are from rotational transitions of CO at high excitation (J = 7
$\rightarrow$ 6), (J = 6 $\rightarrow$ 5) [1] and (J = 3
$\rightarrow$ 2) [2]. This experimental result is extremely
interesting, since it has already been shown that atomic hydrogen
Lyman transitions from quasars can bound the departure from the
known number of space-time dimensions ($\epsilon = D_{{\rm
spacetime}}-4$ = D - 3) in the early universe [3]. The interest on
extra dimensions comes from the fact that they appear naturally in
string theories [4]. In general, there is no known reason why the
dimension of space must have the value of 3, or even be an
integer. Since at present this number is three, and since in the
very early universe there exists the possibility of a larger
number, this indicates that it should be possible to define an
effective number of space dimensions which changes continuously
with time, from its very early universe value to the present one.
Moreover, by looking at relics of the early universe, such as the
cosmic microwave background or the light emitted by quasars with a
very high redshift, it should be possible to measure deviations of
the number of spatial dimensions, $D$, from its present (epoch)
value of 3.

All that is known is that
the present epoch value of $\epsilon$ has an exceedingly small
upper bound [5]. It is an open experimental question whether
$<\epsilon> = 0$ in the early universe, as is apparent in the
current epoch. In reference [3], it was shown that quasar hydrogen
Ly$_{\alpha}$ and Ly$_{\beta}$ carry information about $D =
3+\epsilon$ dimension of space, and using available quasar
database information, a possible positive experimental signal
$\epsilon \neq 0$ was found for $Z_{c} > 4$. A major difference
between this present paper and [3] is that other alternative
spectroscopic signals are utilized here. The Lyman lines past
Ly$_{\alpha}$: (Ly$_{\beta}$, Ly$_{\gamma}, \cdots$) are subject
to absorption by intervening hydrogen clouds (Lyman forest),
leaving in some cases, imprecise locations of the lines. Due to
absorption, the Lyman series in hydrogen becomes difficult to use
for $<\epsilon>$ determination for very high $Z_{c}\geq 4$.
However, rotational spectra of linear molecules, already starting
out in the microwave region, are redshifted towards the radio, so
therefore they do not suffer the intense random absorption events
as optical lines.  We show that they offer unique opportunities
for precise determination of $D = 3+\epsilon$ dimension of space.

The spacing of atomic and molecular energy levels varies with the
dimension of space.
In particular, $\epsilon$ differences between two
fractal geometries give rise to corresponding $\epsilon$
differences in their quantum mechanical energies.

It is, in principle, possible to solve the Schr\"{o}dinger equation in $D =
3+\epsilon$ dimension of space ($4 + \epsilon$ dimension of
space-time), using a Taylor expansion about $D = 3$ [3].

\begin{equation}
<|E|>|_{D \; = \; 3+ \epsilon} =<|E|>|_{D \; = \;3} + \frac{d <|E|> }{d D}|_{D \; = \; 3} \; \epsilon +
\cdots
\end{equation}

The Hellmann-Feynman theorem is

\begin{equation}
\frac{d <|E|> }{d D}|_{D \; = \; 3} \; = \; <| \frac{\partial H}{\partial D}|_{D \; = \; 3} |>
\end{equation}
where $H$ is the D-dimensional Hamiltonian. It can be shown that a
first-order Taylor expansion is always available because physical
energies are continuous at $D = 3$. Because of mathematical
complexity, however, only the hydrogen atom has been presently
amenable to computations of energy levels for $\epsilon \neq 0$.
We now show that another physical system, linear molecules, can
have their energy levels computed for $\epsilon \neq 0$. We are
especially interested in linear molecules, such as CO, because,
being the simplest molecules, they are the ones most likely to be
identified in giant molecular clouds.

The Hamiltonian is

\begin{equation}
H_{r} = B(L)L^{2}
\end{equation}
where $L$ is the body (molecule)-fixed rotational angular momentum
and $B$ is the principal rotational constant (equal to
$\frac{1}{2I}$, where $I$ is the principal moment of inertia).
Because of centrifugal stretching, and higher order effects, $B =
B(L)$. We need to generalize this energy to $D = 3 + \epsilon$
fractal space. In quantum mechanics, $L^{2}$ is a second order
Casimir invariant operator, $C_{2}$. The generalized rotation
operator [6] that correctly incorporates higher order effects is

\begin{equation}
{\cal H}_{r} = B(C_{2})C_{2}
\end{equation}

\begin{equation}
B(C_{2}) = B_{0} + B_{1}C_{2}
\end{equation}
where $B_{0}$ and $B_{1}$ are constants for zero vibrational
quanta states (as is the astrophysical situation). In general, the
second order Casimir invariant is [7]:
\begin{equation}
C_2 = f^i_{jk} f^j_{il} X^k X^l = H_i G_{ij} H_j + \sum_{\rm all\
roots} E^{\alpha} E_{-\alpha}
\end{equation}
where $f^i_{jk}$ are the structure constants, $X^k$ are
generators, and $C_2$ commutes with all generators. The Racah
formula for the eigenvalue of $C_2$ for any irreducible
representation is easily derived by letting $C_2$ act on the state
with highest weight $\Lambda$. The result is:
\begin{equation}
C_2 = (\Lambda,\Lambda+2\delta)\label{e1}
\end{equation}
where $\delta=(1,1,....,1)$ in the Dynkin basis. The scalar
product of any two weights can be written as:
\begin{equation}
(\Lambda,\Lambda') = \sum_{ij} a'_j G_{ij} a _j \label{e2}
\end{equation}
where $G_{ij}$ is a symmetric tensor whose elements can be
computed  for each simple group, and which are given in Table 7 of
reference [7], and the $a_i$ are the Dynkin components of
$\Lambda$.

In our case we want to find the Casimir invariant for the $(L,0,
...,0)$ totally symmetric representation. The choice of
representation depends, of course, on the way in which we want to
generalize angular momentum, and the correct choice, we argue,
would preserve the symmetry properties (in this case this would
mean to keep the completely symmetric coupling) when generalizing
to larger dimensions.

For odd space dimensions $D=2n+1, n=1,2,...$, the algebra is
$B_n$, and for even space dimensions $D  = 2n, n=1,2,...$, it is
$D_n$. We can calculate $C_2$ using equations (\ref{e1}) and
(\ref{e2}), and the $a_j$ values given by $(L,0,...,0)$:
\begin{equation}
C_2 = (L,0,...,0) G(B_n\ {\rm or}\ D_n) (2+L,2,2,...,2)
\end{equation}
which gives, by simple matrix multiplication, both for $B_n$ and
$D_n$:
\begin{equation}
C_2 = L (L+D-2) \  .
\end{equation}
The above formulae can be analytically continued to any
non-integer number of spatial dimensions.

The pure rotational energies of linear molecules
in $D$-dimension space are then:

\begin{equation}
{\cal H}_{rot} = [B_{0} + B_{1}(L(L+1))]L(L+1) + \{2B_{1}L^{2}(L+1) + B_{0}L \} \epsilon
\end{equation}
It is a simple matter to determine $B_{0}=57.635968 \; {\rm GHz}$ and $B_{1}=-1.835 \times 10^{-4} \; {\rm GHz}$ for CO,
giving Table 1.
     \begin{table}
     \begin{tabular}{ccc}
     \multicolumn{3}{c}{Rotational rest-frame $\epsilon$ values} \\  \hline
       Transition & in GHz & in mm\\
       \mbox{} & \mbox{} & \mbox{}\\
   $ 7 \rightarrow 6$  &  806.651719 +57.584588$\epsilon$  & .371650429-0.026531074 $\epsilon$ \\
  $ 6 \rightarrow 5$ & 691.47309 +57.598534$\epsilon$ & .433556218-0.036114496 $\epsilon$ \\
  $3 \rightarrow 2 $ &  345.795991 +57.62716$\epsilon$ & .86696337-0.144480092 $\epsilon$
      \end{tabular}
      \caption{Pure $D$-dimensional rotational levels for CO}
      \end{table}

If $\epsilon \neq 0$, the result is a change in each transition by a
unique amount. The effect is unmistakable: even a very tiny
$\epsilon$ will be easily detectable.

We now evaluate the experimental data [1,2]. The theoretical procedure to determine $\epsilon$ and the error $\delta \epsilon$ is given in [3].
Once two rotational lines are identified, then $\epsilon$ is
derivable by
\begin{equation}
\epsilon = \frac{\tau_{M} \lambda_{1} - \lambda_{0}}
          {a - b \tau_{M}}
\end{equation}
where $$\tau_{M} = \frac{\lambda_{0M}}{\lambda_{1M}}$$ with $\lambda_{0M}$,
$\lambda_{1M}$ the two measured redshifted rotational lines, which
have molecular cloud rest-frame wavelengths $\lambda_{0} + a\epsilon$,
$\lambda_{1}+b\epsilon$, where $\lambda_{0}$, $\lambda_{1}$ are the
present epoch (i.e. laboratory) transition wavelengths. In this
equation, $a$ and $b$ are read off from Table 1.

 Note that here, the error in the determination of the coefficients that multiply $\epsilon$ (labeled $\delta a$, $\delta b$ in [3] ) is zero, because the generalized pure rotation operator has already included the higher order terms.

The instrumental precision centered on the observed frequencies is 50 Mhz for (J = 3 $\rightarrow$ 2) [2] and 5 MHz for both (J = 7 $\rightarrow$ 6), (J = 6 $\rightarrow$ 5) [1]. The observed widths of the lines in km/s is 279 for both (J = 7 $\rightarrow$ 6), (J = 6 $\rightarrow$ 5) and 320 (50 MHz) for (J = 3 $\rightarrow$ 2). We reduce the data to Table 2, using the speed of light c = $2.99792458 \times 10^{11}$ mm/s.

     \begin{table}
     \begin{tabular}{cccc}
     \multicolumn{4}{c}{Processed data} \\  \hline
     Transition  & $\lambda^{obs}$ (mm) &  $\Delta \lambda^{obs}$ (mm) & $\Delta \lambda^{resol}$ (mm) \\
     \mbox{} & \mbox{} &   \mbox{} & \mbox{} \\
      $ 7 \rightarrow 6$ &  108.729(9)/c & 2.56601(-3) & 1.26794(-4) \\
     $ 6 \rightarrow 5$ &  93.204(9)/c & 2.99343(-3) & 1.725526(-4) \\
     $3 \rightarrow 2$ &  46.610(9)/c & 6.865475(-3) & 6.865475(-3)
     \end{tabular}
     \caption{CO lines with experimental precision ($\Delta$). The number in parenthesis is the exponent, i.e. (-3) $\equiv 10^{-3}$}
     \end{table}

The procedure in [3] uses pairs of lines from the same source, so the three lines here yields the unique case of three separate determinations of $\epsilon$ and its error $\delta \epsilon$.
In Table 3, we give the results for the three possible line combinations.

     \begin{table}
     \begin{tabular}{ccc}
     \multicolumn{3}{c}{Early Universe Extra Dimensions} \\  \hline
    microwave pair & $\epsilon$ & $\delta \epsilon$ \\
    \mbox{} &   \mbox{} & \mbox{} \\
       (J = 7 $\rightarrow$ 6) with  (J = 6 $\rightarrow$ 5)  & -0.000012043  & 0.118937801 \\
      (J = 7 $\rightarrow$ 6) with  (J = 3 $\rightarrow$ 2) & -0.000004379 & 0.084521824 \\
      (J = 6 $\rightarrow$ 5) with  (J = 3 $\rightarrow$ 2) & -0.000003283 & 0.096631339
      \end{tabular}
      \caption{$\epsilon$ and error $\delta \epsilon$}
      \end{table}

The average values are

    \begin{eqnarray}
    < \epsilon > &  = &  -0.00000657 \\
    < \delta \epsilon > & = & 0.10003032
    \end{eqnarray}

Taking $< \delta \epsilon> $ as the standard deviation, one can perform the statistical Z-test [8].
This predicts that the probability of $\epsilon \neq 0$ to be one in 7794.

The work of I. S. is supported in part by Fondecyt (Chile) grant
1030355.

\clearpage

\end{document}